\documentclass[12pt]{article}
\usepackage[T1]{fontenc}
\usepackage[utf8]{inputenc}
\usepackage{authblk} 
\usepackage{geometry}
\usepackage{graphicx} 
\usepackage{color}
\usepackage{hyperref}
\usepackage{natbib}
\usepackage{siunitx}
\usepackage{multicol}
\usepackage{listings}
\usepackage{xcolor}

\title{Traditional statistical representations outperform generative AI in identifying expert peer reviewers}

\author[1]{Vicente Amado Olivo\thanks{Correspondence to: amadovic@msu.edu}}
\author[2]{Tereza Jerabkova}
\author[3]{Jakub Klencki}
\author[4]{John Carpenter}
\author[5]{Mario Malički}
\author[6]{Ferdinando Patat}
\author[7]{Louis-Gregory Strolger}
\author[1,8]{Wolfgang Kerzendorf}

\affil[1]{Department of Computational Mathematics, Science, and Engineering, Michigan State University, East Lansing, MI}
\affil[2]{Department of Theoretical Physics and Astrophysics, Faculty of Science, Masaryk University, Kotláˇrská 2, Brno 611 37, Czech Republic}
\affil[3]{Max Planck Institute for Astrophysics, Garching bei München, Germany}
\affil[4]{Joint ALMA Observatory, Alonso de C\'ordova 3107, Vitacura, Santiago, Chile}
\affil[5]{Stanford Program on Research Rigor and Reproducibility (SPORR), Stanford University, CA, USA}
\affil[6]{European Southern Observatory, K. Schwarzschildstr. 2, D-85748, Garching bei M\"unchen, Germany}
\affil[7]{Space Telescope Science Institute, 3700 San Martin Drive, Baltimore, MD 21218, USA}
\affil[8]{Department of Physics and Astronomy, Michigan State University, East Lansing, MI 48824, USA}

\date{}

\begin{document}

\maketitle

\vspace{-2em} 

\vspace{1em}


\begin{abstract}
The exponential growth of scientific submissions has strained the peer review system. Despite the rapidly expanding global pool of researchers, this unprecedented scale has rendered the previous approach of manual expert identification unfeasible. Therefore, institutions have naturally turned to Large Language Models (LLMs) to automate intricate processes like expert reviewer identification. However, the reliability of these new models in accurately identifying domain experts lacks rigorous evaluation. We conduct a comprehensive empirical evaluation of statistical and AI-driven expertise identification methodologies to benchmark their reliability and limitations. Framing expert identification as an information retrieval problem, we utilize the distributed peer review system of a major international astronomical observatory, where proposal authorship serves as our proxy ground truth for domain expertise. Evaluating six retrieval methodologies utilized across observatories and computer science conferences, we demonstrate that traditional statistical representations outperform generative AI. Specifically, Term Frequency-Inverse Document Frequency successfully identified a labeled expert within the top \num{25} recommendations 79.5\% of the time, compared to 51.5\% for \textsc{GPT-4o mini}. Our results highlight that distinguishing subfield expertise requires fine-grained vocabulary, which is obscured by the semantic smoothing in generative methods. By establishing a rigorous evaluation framework for automated peer review, we demonstrate that transparent and reproducible statistical representations still outperform computationally expensive LLMs in specialized scientific tasks.
\end{abstract}



\begin{multicols}{2}

\section{Introduction}\label{sec:intro}
The exponential growth of scientific outputs is placing unprecedented strain on the global peer review system, challenging the sustainability of research publishing, grant funding, and resource allocation \citep{Kovanis2016TheGB, Hanson2023TheSO}. As submissions surge across all disciplines, the reliance on manual workflows to identify and assign qualified experts for peer review has become a bottleneck \citep{Zhao2021ReviewerRU, Saveski2023CounterfactualEO,Xu2026ArtificialIA}. In response, institutions, such as astronomical observatories and computer science conferences, are rapidly adopting automated systems that leverage Large Language Models (LLMs) and statistical methods to estimate scientific expertise and match reviewers to submissions \citep{charlin_framework_2012, kerzendorf_distributed_2020, Lee2025TheRO, stelmakh_gold_2025, Teixeira2025AIIP}. However, integrating automated systems into the scientific process requires more than just administrative efficiency; it demands rigorous evaluation against the core scientific standards of transparency, reproducibility, and robust expert identification.

Structurally, automating reviewer identification and assignment is a two-stage process: first, generating similarity scores to estimate expertise, and second, computationally optimizing the final assignments subject to constraints (e.g., conflicts of interest, reviewer workload) \citep{stelmakh_peerreview4all_2019,leyton-brown_matching_2022}. To generate these scores, automated methodologies generally  map both reviewer bibliographies and research output texts (e.g., publications or proposals) into shared, high-dimensional vector spaces to calculate similarity \citep{mimno_expertise_2007, charlin_framework_2012, zhang_multi-label_2019}. Without an objective definition of a scientific 'expert,' current methods rely on topical alignment as the primary signal of expertise, delegating qualifications (e.g., seniority) to the constraint-based optimization \citep{kerzendorf_distributed_2020, carpenter_enhancing_2025}. 

To handle these scaling challenges, disciplines from astronomy to computer science have adopted various machine learning methods to automate reviewer identification and assignment \citep{kerzendorf_distributed_2020, stelmakh_gold_2025}. For example, the Space Telescope Science Institute and the Toronto Paper Matching System both utilize Term Frequency-Inverse Document Frequency (TF-IDF) to align reviewers to proposals or papers\citep{strolger_proposal_2017, strolger_pacman2_2023, charlin_framework_2012}. Similarly, the Atacama Large Millimeter Array and several computer science pipelines infer expertise using topic modeling via Latent Dirichlet Allocation (LDA) \citep{meyer_analysis_2022, carpenter_enhancing_2025, rosenzvi2012authortopicmodelauthorsdocuments, mimno_expertise_2007, Anjum2019PaReAP}. More recently, conference platforms like OpenReview have deployed Transformer-based semantic embeddings \citep{noauthor_openreviewopenreview-expertise_2025}, while funding bodies like the European Research Executive Agency are exploring specialized LLMs to match experts to fellowship proposals \citep{lvarezGarca2026ExpertAS}. These semantic approaches contrast with the explicit keyword-based matching currently utilized by the European Southern Observatory \citep{jerabkova_first_2023}. Ultimately, the fragmented landscape of varied algorithms highlights a critical lack of consensus, underscoring the urgent need to systematically evaluate how these representations retrieve experts in operational environments \citep{stelmakh_gold_2025}.

Despite the widespread deployment of these methodologies, systematically benchmarking their efficacy is hindered by the absence of ground-truth labels for scientific expertise, forcing prior evaluations to rely on inherently biased proxy labels \citep{Anjum2019PaReAP, noauthor_openreviewopenreview-expertise_2025, stelmakh_gold_2025}. For instance, external human annotation introduces noise as annotators may lack specific reviewer knowledge \citep{mimno_expertise_2007, Zhao2021ReviewerRU}. Conversely, self-assessments are susceptible to calibration biases (e.g., self-efficacy bias) \citep{Dumais1992AutomatingTA, Kruger1999UnskilledAU, Ehrlinger2003HowCS, Rodriguez_2008, aitymbetov-zorbas-2025-autonomous}, and proxying expertise strictly through authorship \footnote{See OpenReview Github repository: \url{https://github.com/openreview/openreview-expertise}} \citep{noauthor_openreviewopenreview-expertise_2025} may not capture a reviewer's broader domain competency \citep{stelmakh_gold_2025}. Ultimately, the field lacks large-scale, systematic evaluations of the underlying representation methodologies. 

Major astronomical observatories provide a unique environment to systematically benchmark expertise identification methods. The allocation of telescope time represents a competitive resource distribution system, demanding precision in expert evaluation \citep{strolger_proposal_2017, kerzendorf_distributed_2020}. Crucially, the recent adoption of Distributed Peer Review (DPR) frameworks at various facilities, where proposal authors simultaneously act as the reviewer pool, creates a closed-loop system that allows for the rigorous, large-scale evaluation of expertise identification methods \citep{merrifield_telescope_2009, 2019Msngr.177....3P}. Capitalizing on the DPR system deployed by the European Southern Observatory (ESO), we introduce a dual validation strategy using two distinct measures: first, we treat a researcher’s submitted proposal as a 'proxy label' for their expertise; second, we compare these results against the expertise categories researchers selected for themselves (self-reported labels). Treating expert identification as an Information Retrieval (IR) problem, we evaluate six distinct methodologies utilized across astronomy and computer science: keywords, TF-IDF, LDA, two Transformer embedding models, and \textsc{GPT-4o mini}. Using this operational dataset from ESO's Period~110 (P110) call, we conduct a large-scale benchmark study across \num{379} reviewers and \num{435} proposals to determine how effectively these representations retrieve true domain experts.


\section{Data}\label{sec:data}
The data for this study originate from the P110 observing call at ESO. P110 refers to the six-month scheduling cycle for ESO telescope time that began in late 2022 \citep{jerabkova_first_2023}. During this cycle, ESO first deployed the DPR system for all proposals requesting fewer than \num{16} hours of observing time. In total, P110 received \num{435} eligible proposals involving \num{2014} unique investigators. Under the DPR framework, submitting teams are required to nominate one investigator to serve as a reviewer. Throughout this paper, we refer to this individual as the “proposal-designated reviewer” to distinguish them from other co-investigators. Because some individuals were nominated to represent multiple proposals, the \num{435} submissions resulted in a final pool of \num{379} unique reviewers. Each proposal-designated reviewer was assigned to assess \num{10} proposals. 

The primary data for this study consist of (1) the proposal text and (2) reviewer profiles (e.g., self-reported keywords or publication abstracts). Each proposal includes multiple sections, from scientific justification to feasibility, but our experiments focus on the proposal abstract. This serves as a concise summary of the scientific content, with the longest proposal abstract in P110 containing \num{176} words. We rely on two data sources for representing reviewer expertise: ESO keyword data and publication abstracts retrieved from the NASA/ADS digital library, which offers an open API. Using the ADS Python package\footnote{\url{https://ads.readthedocs.io/en/latest/}}, we query abstracts for each proposal-designated reviewer by name. Alongside the keywords and abstracts, we utilize the self-reported expertise labels collected during the P110 DPR at ESO. For each of the \num{4350} reviewer–proposal assignments (corresponding to the \num{435} proposals, each assigned to ten reviewers), the assigned reviewer categorized their own expertise into one of three distinct labels: Expert, Intermediate, or Non-expert. Detailed review statistics, NASA/ADS query descriptions, and the baseline ESO keyword distributions are provided in the SI Appendix~\ref{sec:data_SI}.

\section{Methods}\label{sec:methods}
To evaluate how effectively each approach estimates reviewer expertise, we first formalize the task of identifying an expert reviewer as an information retrieval problem before describing our experimental setup and detailing the individual representation methods in the SI Appendix~\ref{sec:methods_SI}.

\subsection{Problem Formulation}
We frame the task of identifying expert reviewers as an information retrieval problem. In this formulation, each proposal acts as a query seeking to return the most relevant results, which corresponds to a set of reviewers. The goal of each representation method is to rank reviewers from most suitable to least for a given proposal.

In our study $p_i$ denotes the embedding representation of proposal $i$, and $r_j$ denotes the embedding representation of reviewer $j$ as produced by a given method. A similarity function $d(p_i, r_j) = s_{ij}$, such as cosine similarity, assigns a similarity score between the proposal and reviewer vector representations.

For each method, we first compute vector representations for each unique reviewer's publication history ($N = \num{379}$) and for each proposal abstract ($N = \num{435}$). We normalize all vectors using the $L_2$ norm to project them onto the unit hypersphere. By normalizing the vectors, we measure only the direction of the expertise, ignoring the length (or word count) of a reviewers collection of abstracts. We then calculate pairwise cosine similarities between all reviewer and proposal vectors, producing an expertise matrix that represents the similarity between each reviewer’s expertise and every proposal. The resulting matrix has dimensions ~\num{435}$\times$\num{379}~, where each entry $s_{ij}$ denotes the similarity between proposal $p_i$ and reviewer $r_j$. The IR formulation provides the foundation for our evaluation framework, where we assess how effectively each representation ranks reviewers according to their represented expertise.

\begin{figure*}[t!]
    \includegraphics[width=\textwidth]{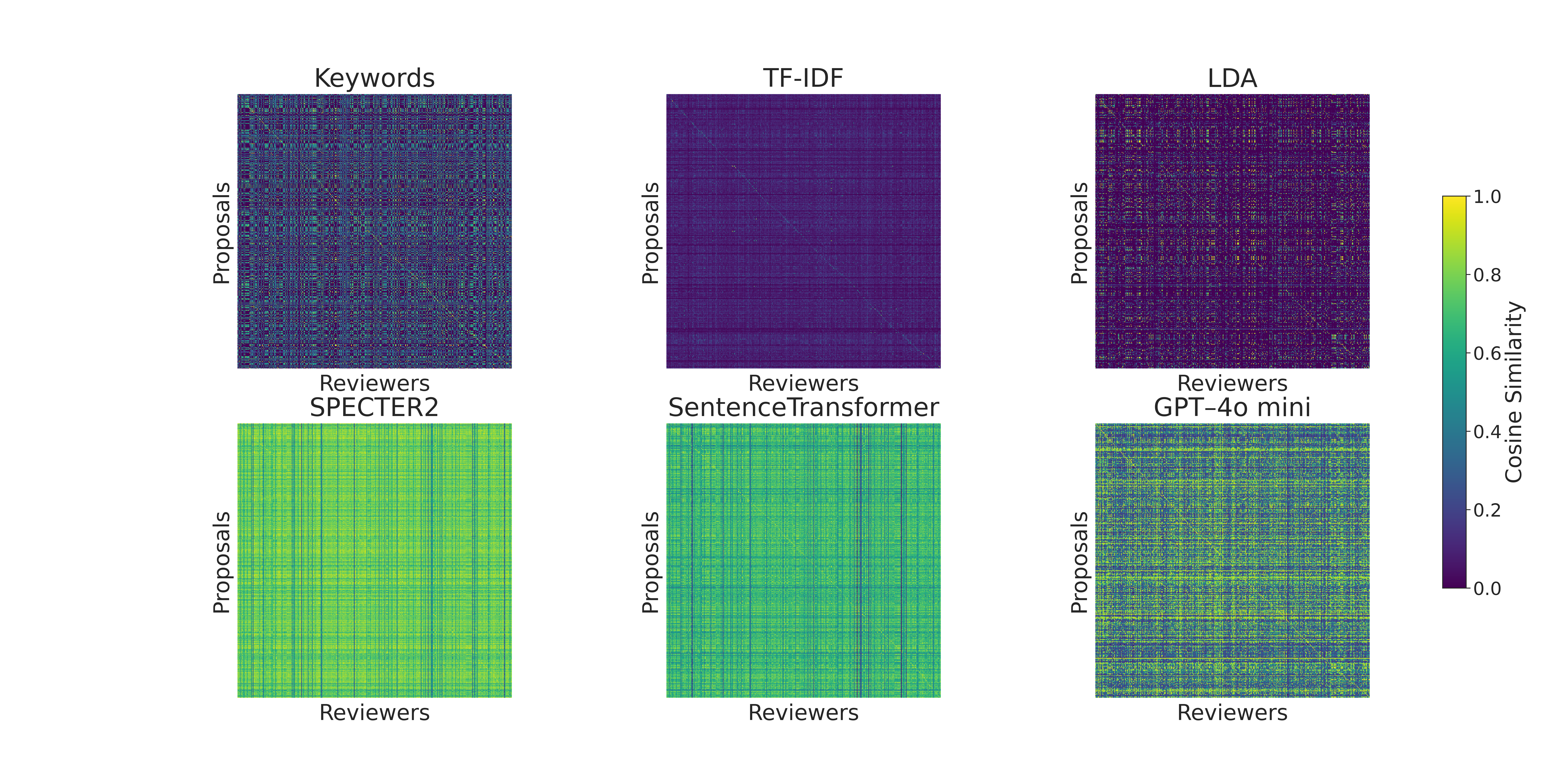}
    \caption{Similarity score matrices across methods for Period P110. Rows and columns are sorted such that the proposal-designated reviewer-proposal pairs align along the diagonal.}
    \label{fig:score_matrices}
\end{figure*}

\subsection{Experimental Setup}\label{subsec:metrics}
To systematically compare the various expertise representation methods, we established a standardized evaluation framework. Although each method maps textual inputs into a continuous vector space, these spaces differ significantly in scale and distribution, rendering raw similarity scores directly incomparable. Therefore, our evaluation relies on rank-based IR metrics to assess the ordering of reviewers based on similarity scores. In the absence of a standardized ground-truth benchmark for expertise modeling \citep{stelmakh_gold_2025}, we adopt a dual validation strategy. We evaluate each methodology against two distinct sources of labels derived from the ESO DPR system: (1) proxy gold labels based on proposal authorship and (2) self-reported reviewer expertise.

\begin{table*}[t!]
\centering
\caption{Average rank-based retrieval performance across 435 proposals. Values are reported as mean $\pm$ margin of error (95\% CI). Statistically significant differences from the baseline (Keywords) are marked with * ($p < 0.05$) and \dag ($p < 0.01$). \textbf{Bold} indicates the best performance.}
\resizebox{\textwidth}{!}{
\begin{tabular}{lcccc}
\hline
Method & Median Rank $\downarrow$ & MRR $\uparrow$ & Hit@25 $\uparrow$ & Z-score$\uparrow$ \\
\hline
Keywords & 9.0 \scriptsize{$\pm 2.5$} & 0.270 \scriptsize{$\pm 0.032$} & 0.703 \scriptsize{$\pm 0.044$} & 2.051 \scriptsize{$\pm 0.095$} \\
LDA (K = 50) & 24.0 \scriptsize{$\pm 6.0$ \dag} & 0.148 \scriptsize{$\pm 0.024$ \dag} & 0.503 \scriptsize{$\pm 0.047$ \dag} & 2.114 \scriptsize{$\pm 0.208$} \\
TF--IDF & \textbf{4.0} \scriptsize{$\pm 1.0$ \dag} & \textbf{0.408} \scriptsize{$\pm 0.037$ \dag} & \textbf{0.795} \scriptsize{$\pm 0.038$ \dag} & \textbf{3.969} \scriptsize{$\pm 0.340$ \dag} \\
SPECTER2 (Mean pooling) & 10.0 \scriptsize{$\pm 2.5$ \dag} & 0.288 \scriptsize{$\pm 0.034$} & 0.634 \scriptsize{$\pm 0.046$ *} & 0.834 \scriptsize{$\pm 0.114$ \dag} \\
SentenceTransformer(Mean pooling) & 7.0 \scriptsize{$\pm 2.0$} & 0.341 \scriptsize{$\pm 0.036$ \dag} & 0.710 \scriptsize{$\pm 0.043$} & 1.252 \scriptsize{$\pm 0.123$ \dag} \\ 
GPT--4o mini & 24.0 \scriptsize{$\pm 4.0$ \dag} & 0.166 \scriptsize{$\pm 0.027$ \dag} & 0.515 \scriptsize{$\pm 0.047$ \dag} & 1.684 \scriptsize{$\pm 0.079$} \dag \\
\hline
\end{tabular}
}
\label{tab:retrieval_results}
\end{table*}

First, we utilize the unique structure of DPR to assign proxy gold labels. As each proposal designates an investigator to participate as a reviewer in the DPR process, we define each proposal-designated reviewer as a proxy gold expert on their own proposal. This formulation enables each proposal–reviewer ranking to be evaluated as an IR task. To quantify the quality of the vector representations output by each expertise modeling technique, we compute four evaluation metrics for each proposal-designated reviewer. First, we report the Median Rank of the proposal-designated reviewer within the sorted list of all candidate reviewers. Second, we compute the Mean Reciprocal Rank (MRR), defined as the average of the inverse of the rank ($\frac{1}{\text{rank}}$) across all proposals, which penalizes lower rankings more heavily. Third, we measure Hit@25, the fraction of proposals where the proposal-designated reviewer appears within the top 25 candidates. Finally, to assess how significantly the designated reviewer stands out from the entire sample of reviewers, we calculate a standardized $z$-score based on the distribution of similarity scores. For a given proposal, this is defined as:$$z = \frac{S_{\text{I}} - \mu}{\sigma}$$where $S_{\text{I}}$ is the similarity score assigned to the proposal-designated reviewer, and $\mu$ and $\sigma$ are the mean and standard deviation, respectively, of the similarity scores for all potential reviewers for that specific proposal.

Finally, we benchmark the expertise representations against self-reported expertise labels from P110. We utilize the Normalized Discounted Cumulative Gain (NDCG) metric for evaluation to account for the hierarchy of self-reported labels, penalizing models that fail to rank an 'Expert' above a 'Non-expert' (see SI Appendix \ref{subsec:NDCG_SI} for the formal formulation). We utilize the scikit-learn implementation of $\text{NDCG}$ \citep{pedregosa_scikit-learn_2018}. In our evaluation, a graded relevance score is assigned based on the reviewer's self-reported expertise for the subset of reviewer-proposal pair assignments in P110 totaling \num{4350} (\num{10} for each proposal). The relevance scores are weighted to reward the correct ranking of 'Expert' reviewers (assigned a score of 10), while only providing a minor gain for 'Intermediate' reviewers (assigned a score of 2), and no gain for 'Non-Expert' candidates. This evaluation metric emphasizes the importance of ranking labeled 'Experts' above 'Non-experts'.

To ensure that our performance metrics are robust and not artifacts of specific sample selection, we calculate 95\% confidence intervals for all aggregate results. We utilize bootstrap resampling ($n=10,000$, percentile method) \citep{Efron1979BootstrapMA}, which allows us to estimate the uncertainty of our metrics (e.g., the error bars on MRR) without assuming a specific underlying distribution. Furthermore, to rigorously compare the proposed methods against the baseline (i.e., keywords), we employ the Wilcoxon signed-rank test \citep{Wilcoxon1945IndividualCB}. Unlike standard tests that assume Gaussian errors, this non-parametric test is better suited for Information Retrieval metrics like Rank and MRR, which are typically heavily skewed. We report differences as statistically significant when the probability of the result occurring by chance is less than 5\% ($p < 0.05$).

\begin{figure*}[t!]
    \centering
    \includegraphics[width=0.7\textwidth]{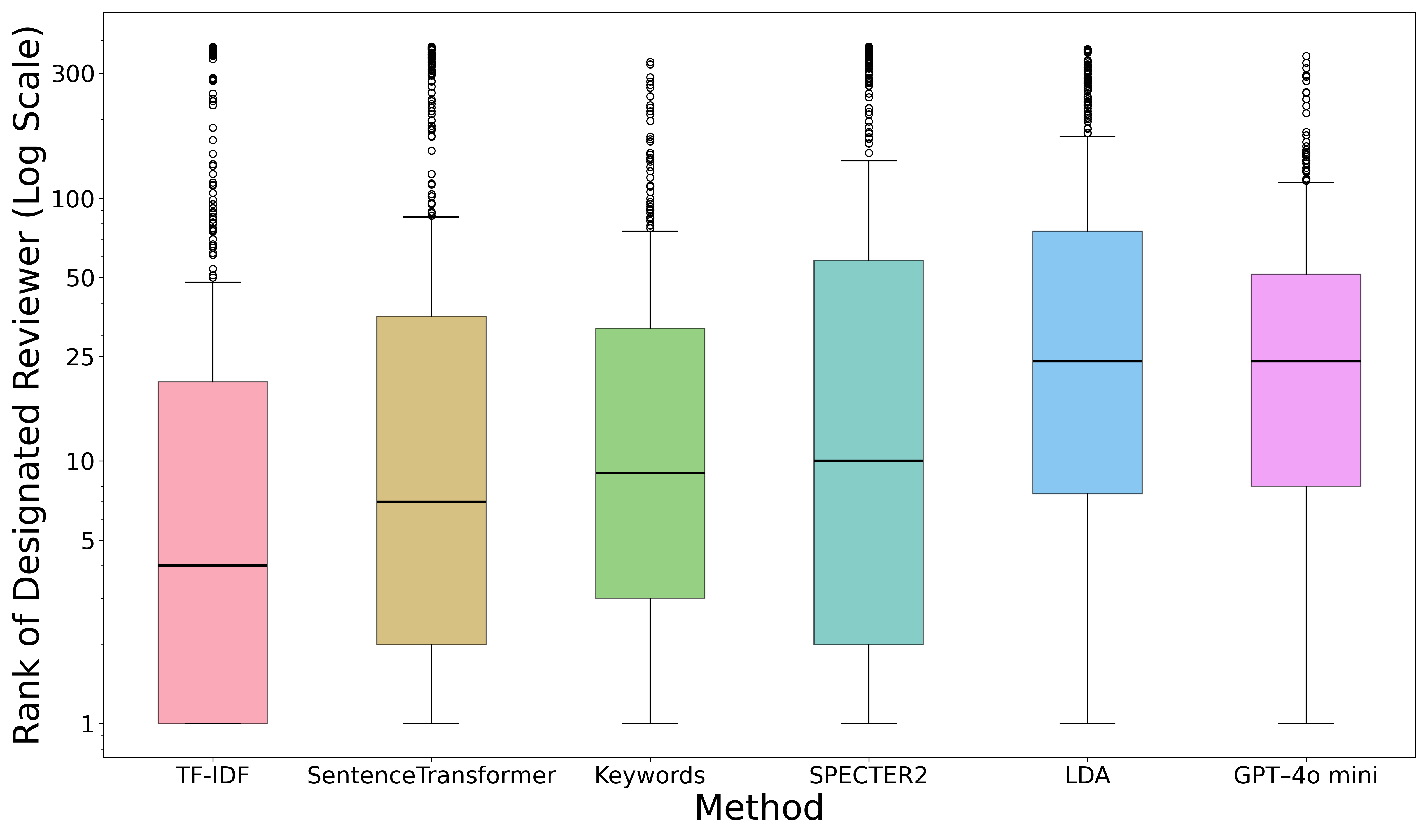}
    \caption{Distribution of the rank assigned to the proposal-designated reviewer across 435 proposals. Boxes span the 25th to 75th percentiles, with whiskers showing the range of non-outlier data. The y-axis is logarithmic to accommodate the wide variance in rankings. TF-IDF consistently retrieves the designated expert at the top of the list (median rank 4), compared to Transformer methods (SPECTER2, SentenceTransformer or GPT-4o mini).}
    \label{fig.score_distributions}
\end{figure*}

\section{Results} \label{sec:results}

Comparing raw scores across methods is difficult because different methods produce very different score distributions and sparsity patterns. Table~\ref{tab:distribution_stats} presents the similarity score distributions across different methods. Keyword-based and topic modeling methods exhibit high sparsity, though for different reasons: keywords produce roughly 52\% zero-valued similarities. For topic modeling, while absolute zero similarities are mathematically rare, approximately 68\% of the scores fall below 0.01. Because standard implementations (e.g., Gensim) default to truncating probabilities below this 0.01 threshold, these practically operate as zero-valued similarities in the matrix. In contrast, TF–IDF produces less than one percent zero scores, but its maximum score is only \num{0.50}. Conversely, the SPECTER2 (Mean pooling) embeddings yield uniformly high similarities with a minimum score of \num{0.75}. Therefore, simply comparing raw score magnitudes alone does not indicate the superiority of one method over another.

Figure~\ref{fig:score_matrices} presents the reviewer-proposal similarity matrices where rows correspond to reviewers and columns to proposals. We order matrices so that the proposal-designated reviewer pairs lie approximately on the diagonal. Because a proposal's designated reviewer is expected to be knowledgeable about their own proposal, a visible diagonal provides a baseline check that each method captures reviewer–proposal similarity alignment. 

\begin{table*}[t!]
    \centering
    \caption{Alignment with human self-reported expertise. We report Normalized Discounted Cumulative Gain (NDCG) to measure ranking quality. Values are reported as mean $\pm$ margin of error (95\% CI). Statistically significant differences from the baseline (Keywords) are marked with * ($p < 0.05$) and \dag ($p < 0.01$). \textbf{Bold} indicates the best performance.}
    \label{tab:human_alignment}
    \begin{tabular}{lc}
    \hline
        Method & NDCG $\uparrow$ \\
        \hline
        TF-IDF & \textbf{0.832} \scriptsize{$\pm 0.014$ *} \\
        SentenceTransformer (Mean pooling) & 0.827 \scriptsize{$\pm 0.014$ *}\\
        SPECTER2 (Mean pooling) & 0.815 \scriptsize{$\pm 0.014$ *}\\
        GPT--4o mini & 0.811 \scriptsize{$\pm 0.011$ *}\\
        LDA & 0.798 \scriptsize{$\pm 0.015$} \\
        Keywords & 0.789 \scriptsize{$\pm 0.014$} \\
        \hline
    \end{tabular}
\end{table*}

Therefore, we utilize standard retrieval-based metrics to quantitatively compare the performance of each method. We evaluate the similarity scores for each method in two stages: first, by measuring the ability of each method to retrieve the proposal-designated reviewer as a known proxy label of expertise, and second, by assessing how the computed similarity scores align with the reviewers' self-reported expertise for assigned matches. Table \ref{tab:retrieval_results} details the rank-based retrieval performance across the \num{435} proposals. We report values with 95\% confidence intervals to account for sample variance and determine statistical significance relative to the Keyword baseline. As shown in Figure \ref{fig.score_distributions}, TF–IDF demonstrates the most robust retrieval performance across all evaluated methods, consistently placing the designated reviewer higher in the ranked list. It identified the proposal-designated reviewer with a Median Rank of 4.0 ($\pm 1.0$) and a Hit@25 of 0.795 ($\pm 0.038$). The SentenceTransformer (Mean pooling) method follows with a Median Rank of 7.0 ($\pm 2.0$) and a Hit@25 of 0.710 ($\pm 0.043$). The Keyword baseline yields a Median Rank of 9.0 ($\pm 2.5$) and a Hit@25 of 0.703 ($\pm 0.044$). Notably, while the Keyword method retrieves the designated reviewer at a similar rate to the SentenceTransformer, it achieves a lower MRR of 0.270 compared to 0.341, suggesting the Transformer consistently ranks the designated reviewer higher within that top 25. Conversely, SPECTER2 (Mean pooling) yields a higher Median Rank of 10.0 ($\pm 2.5$) and underperforms the baseline in recall, achieving a Hit@25 of only 0.634 ($\pm 0.046$). Finally, both LDA and \textsc{GPT-4o mini} underperform the baseline ($p < 0.01$). \textsc{GPT-4o mini} yields a Median Rank of 24.0 ($\pm 4.0$) and a Hit@25 of 0.515 ($\pm 0.047$), performing comparably to LDA, which achieved a Median Rank of 24.0 ($\pm 6.0$) and a Hit@25 of 0.503 ($\pm 0.047$).

Complementing the proposal-designated reviewer retrieval evaluation, we examine how the computed similarity scores align with reviewers’ self-reported expertise labels for their assigned proposals. An effective expertise representation should yield consistently higher similarity scores for self-identified 'Experts' relative to 'Non-Experts.' Table~\ref{tab:human_alignment} presents the NDCG across methodologies. Notably, the automated text representations outperform the self-reported Keywords baseline in aligning with human judgment. TF-IDF achieves the highest alignment with an NDCG of 0.832 ($\pm 0.014$), closely followed by the SentenceTransformer (Mean pooling) at 0.827 ($\pm 0.014$). Both methods demonstrate a statistically significant improvement over the baseline ($p < 0.05$). SPECTER2 (Mean pooling) and GPT-4o mini also perform strongly, yielding NDCG scores of 0.815 ($\pm 0.014$) and 0.811 ($\pm 0.011$), respectively, and both significantly outperform the baseline. Conversely, the explicitly selected Keywords baseline exhibits the weakest alignment (0.789 $\pm 0.014$), performing comparably only to LDA (0.798 $\pm 0.015$), which is the only automated method that does not show a statistically significant improvement over the Keywords baseline.

\section{Discussion}\label{sec:discussion}
Our results demonstrate that traditional statistical lexical methods (i.e., TF-IDF) outperformed both semantic neural architectures and generative models in retrieving proposal-designated reviewers. By leveraging exact matches of domain-specific vocabulary (e.g., \textit{Type Ia}, \textit{Brown Dwarf}), TF-IDF encodes precise expertise signals. In contrast, the latent semantic clustering of neural embeddings (SentenceTransformer and SPECTER2) applies a smoothing effect that obscures the fine-grained distinctions required to differentiate highly specialized sub-domain expertise. Although most methods effectively identified relevant experts, \textsc{GPT-4o mini} and LDA significantly underperformed the Keywords baseline in median rank and recall ($p < 0.01$). Initial exploratory tests with the latest generation of reasoning-based GPT models yielded retrieval performance similar to \textsc{GPT-4o mini}. The architecture of newer models currently disables the parameter controls (e.g., temperature) \citep{openai2026openaio1card} necessary to minimize generation variance and maximize reproducibility for formal benchmarking. To ensure scientific reproducibility in our formal evaluation, we utilized automated context retrieval, structured system prompting, and deterministic guardrails (see SI Appendix, Section~\ref{sec:transformer_SI} for details on our harness).

Beyond baseline performance, TF-IDF and SentenceTransformer also proved highly robust in data-constrained scenarios, maintaining stable performance where LDA and SPECTER2 degraded with limited publication histories (see SI Appendix, Tables \ref{tab:ablation_1}--\ref{tab:ablation_3}). Furthermore, algorithmic hyperparameters influenced retrieval accuracy; Max pooling severely degraded Transformer performance (see SI Appendix, Table \ref{tab:maxpooling}), and LDA exhibited sensitivity to the configured topic count (see SI Appendix, Table \ref{tab:k_topics}). Additionally, deploying generative LLMs introduces substantial computational costs and relies on probabilistic generation, which undermines the  deterministic reproducibility required for peer review decisions and allocations (see SI Appendix, Section \ref{sec:limitations_SI}). Finally, automated text representations, led by TF-IDF (NDCG = 0.832), better aligned with human self-reported expertise than the self-selected Keywords baseline, demonstrating that abstracts capture a reviewer's domain expertise more comprehensively than coarse manual categorization.

Our findings contrast with recent literature \citep{stelmakh_gold_2025}, which concluded that TF-IDF requires the full text of manuscripts to compete with specialized deep-learning models like SPECTER2. In our evaluation, TF-IDF outperforms semantic neural architectures even when all methods are strictly constrained to publication abstracts. We attribute this divergence to fundamental differences in dataset composition and the framing of the evaluation task. \citet{stelmakh_gold_2025} constructed a dataset of \num{58} researchers evaluating 5 to 10 papers they had previously read, testing algorithms on their ability to correctly order the small, localized set of papers for a specific researcher. Our evaluation explicitly mirrors the operational reality of proposal/conference peer review administration: globally searching a candidate pool to identify the most relevant experts. 

We acknowledge that this study relies on proxy labels, such as proposal authorship and self-reported confidence, as no objective ground truth for scientific expertise exists \citep{stelmakh_gold_2025}. While proposal authorship is a strong proxy for domain knowledge, it may introduce a bias toward a researcher's vocabulary usage rather than capturing broader comprehension of the field \citep{stelmakh_gold_2025}. Similarly, self-reported expertise labels introduce self-efficacy bias, as they rely on the reviewer's personal perception of their knowledge relative to a proposal rather than a standardized taxonomy \citep{Ehrlinger2003HowCS}. Finally, our evaluation is limited to publication abstracts due to the logistical and copyright barriers associated with obtaining full-text access at scale. A comprehensive discussion of further methodological constraints, including author name disambiguation limitations \citep{olivo2025practicalauthordisambiguationmetadata} and reviewer data completeness, is provided in the SI Appendix~\ref{sec:limitations_SI}.

\section{Conclusion}\label{sec:conclusion}
The sustainability of global peer review hinges on the ability to efficiently and accurately match submissions to true domain experts at scale. By formalizing expert identification as a retrieval problem, our evaluation leveraging ESO's Distributed Peer Review data highlights that traditional statistical representations outperform modern neural and generative architectures. The success of TF-IDF demonstrates that in specialized scientific domains, the matching of precise technical vocabulary is more discriminative than the broad semantic clustering of LLMS. As funding agencies and observatories increasingly automate their peer review systems, our evaluation demonstrates that transparent, computationally efficient lexical methods remain the most robust standard for scientific expertise retrieval.

\paragraph{\textbf{Acknowledgments}}
We acknowledge the data provided by the European Southern Observatory (ESO) regarding Distributed Peer Review (DPR) for Period 110. This research has made use of the NASA Astrophysics Data System (ADS) Bibliographic Services and its API. Additionally, this work was supported by the National Science Foundation Research Traineeship Program (DGE-2152014) for Vicente Amado Olivo. The work of Mario Malički is supported by the Stanford School of Medicine Research Office. We also acknowledge the use of the following software packages: \texttt{scikit-learn} \citep{scikit-learn}, \texttt{gensim} \citep{gensim}, \texttt{transformers} \citep{transformers}, \texttt{openai} \citep{openai}, and \texttt{sentence-transformers} \citep{sentence_transformer}.

\paragraph{Data and Code Availability}
The specific proposal abstracts, reviewer identities, and self-reported expertise labels from the European Southern Observatory Period 110, contains sensitive investigator information and is not publicly available. All code for the expertise representation methodologies (TF-IDF, LDA, Transformer embeddings, and GPT-4o mini scoring) and the information retrieval evaluation framework is available at \url{https://github.com/deepthought-initiative/peer_review_expertise_retrieval}. Because the operational ESO data is restricted, the repository includes functionality to generate a synthetic distributed peer review dataset. This tool programmatically retrieves publicly available proposal authors and abstracts from the Hubble Space Telescope and James Webb Space Telescope via the NASA/ADS API to create a "dummy" benchmark. This framework enables researchers to verify the codebase and benchmark various expertise representations using realistic astronomical literature, ensuring reproducibility without compromising sensitive ESO records.

\paragraph{\textbf{Contributor Roles}}
\begin{enumerate}
    \item Conceptualization: Vicente Amado Olivo, Tereza Jerabkova
    \item Data Curation: Tereza Jerabkova, Vicente Amado Olivo, Jakub Klencki
    \item Formal Analysis: Vicente Amado Olivo, Tereza Jerabkova
    \item Funding Acquisition: Tereza Jerabkova, Wolfgang Kerzendorf
    \item Investigation: Vicente Amado Olivo, Tereza Jerabkova, Jakub Klencki
    \item Methodology: Vicente Amado Olivo, Tereza Jerabkova 
    \item Project Administration: Tereza Jerabkova
    \item Resources: Tereza Jerabkova, Wolfgang Kerzendorf
    \item Software: Vicente Amado Olivo, Jakub Klencki
    \item Supervision: Wolfgang Kerzendorf
    \item Validation: Tereza Jerabkova, Wolfgang Kerzendorf 
    \item Visualization: Vicente Amado Olivo, Jakub Klencki
    \item Writing - original draft: Vicente Amado Olivo
    \item Writing - reviewing \& editing: Vicente Amado Olivo, Tereza Jerabkova, Jakub Klencki, John Carpenter, Mario Malički, Ferdinando Patat, Louis-Gregory Strolger, and Wolfgang Kerzendorf
\end{enumerate}

\end{multicols}{2}

\bibliography{expertise}
\bibliographystyle{abbrvnat}

\clearpage 
\appendix

\renewcommand{\appendixname}{SI Appendix}

\renewcommand{\thesection}{S\arabic{section}}
\renewcommand{\thefigure}{S\arabic{figure}}
\renewcommand{\thetable}{S\arabic{table}}

\renewcommand{\theHsection}{S\arabic{section}}
\renewcommand{\theHfigure}{S\arabic{figure}}
\renewcommand{\theHtable}{S\arabic{table}}

\setcounter{figure}{0}
\setcounter{table}{0}

\section*{Supporting Information (SI) Appendix}

\section{Data Details}\label{sec:data_SI}

\subsection{Reviewer Statistics and Assignment}
As outlined in the main text, the Distributed Peer Review process for the European Southern Observatory Period~110 (P110) utilized a final pool of \num{379} unique proposal-designated reviewers. Over 90\% of these reviewers were the proposal PI, while the remainder were either Co-PIs or designated proxies. Regarding career stage, roughly 15\% of the reviewer pool consisted of students, with the remainder being postdoctoral researchers or professional astronomers. 

\subsection{Baseline Keyword Matching Pipeline}
In P110, the reviewer–proposal matching relied strictly on keyword-based similarity \citep{2025Msngr.194...33J}. Reviewers provided keywords when creating their ESO User Portal profiles, and principal investigators assigned keywords to their submitted proposals. Across all proposals and reviewers in P110, there were \num{138} unique keywords chosen from ESO's standardized corpus\footnote{See the full P110 keyword list at: \url{https://www.eso.org/p1demo/proposals/19859/keywords}}. ESO developed an internal algorithm (detailed in Section \ref{sec:methods_SI}) that transforms these selected keywords into vector representations to calculate similarity scores for all reviewer-proposal pairs. Based on these similarity scores, the system assigns ten experts to each proposal using the \texttt{peerreview4all}\footnote{\url{https://github.com/niharshah/peerreview4all}} optimization algorithm \citep{stelmakh_peerreview4all_2019, 2025Msngr.194...33J}.

\subsection{Publication Data Retrieval}
To construct the text-based expertise profiles for evaluation, we retrieved reviewer publication histories via the NASA Astrophysics Data System API. When querying for a maximum of \num{25} publications over the last five years, the mean number returned per reviewer was \num{22}. Notably, over 68\% of these queries returned the maximum limit of \num{25} publications. Because the API supports fine-grained filtering (e.g., by publication year, authorship position, or publication count), we experimented with multiple query strategies to evaluate how different constraints on a reviewer's publication history affect the resulting vector representations and downstream retrieval performance (see Section~\ref{sec:ablation_results}).

\subsection{ESO Keyword Categories}\label{subsec:eso_categories}
In addition to specific sub-field keywords, the ESO proposal system organizes proposal and reviewer subjects into the following nine broad scientific categories. These categories serve as the high-level classification for all submissions:

\begin{enumerate}
\item Physical data and processes
\item Astrometry and celestial mechanics
\item The Sun
\item Planetary systems
\item Stars
\item Interstellar medium (ISM), nebulae
\item The Galaxy
\item Galaxies
\item Cosmology
\end{enumerate}

\section{Representation Methods}\label{sec:methods_SI}
This section details the mathematical formulations and algorithmic implementations for the expertise representation methodologies evaluated in the main text. We begin by outlining the baseline keyword-matching framework currently deployed by the European Southern Observatory, followed by the probabilistic, statistical, and neural architectures utilized in our benchmark.

\subsection{ESO Keywords}
The ESO DPR system represents both reviewers and proposals as keyword vectors representing self-reported scientific keywords. Each reviewer and proposal specifies between two and five keywords in decreasing order of relevance from a static vocabulary compiled by ESO \footnote{See the keyword list at \url{https://www.eso.org/p1demo/proposals/19859/keywords}} \citep{primas_new_2019}. Reviewer keywords are specified once during the creation of a reviewer’s profile in the ESO portal and can be updated individually, however, it is unknown how frequently reviewers revise them. Reviewers are instructed to rank their keywords by relevance:
\[
W = n_{\text{max}} - \text{order} + 1,
\]
where \(n_{\text{max}} = 5\). Thus, the first keyword receives a weight of 5, the second 4, and so on.

These weights populate a vector corresponding to the global keyword corpus, where all positions are zero except for those representing the selected keywords for a given proposal or reviewer. For example, if a reviewer selects "Stars: supernovae", "Stars: binaries", and "Galaxies: abundances" as their top three keywords, only those three positions in the vector receive nonzero weights (5, 4, and 3 respectively). The similarity between a reviewer (\(p_i\)) and a proposal (\(r_j\)) is then computed as the cosine similarity between the two vectors:
\[
s_k = \frac{p_i \cdot r_j}{|p_i| \, |r_j|},
\]
which outputs a score between 0 (i.e., no overlapping keywords) and 1 (i.e., identical keyword lists in the same order).

If the reviewer and the proposal have no shared keywords, a secondary category-level similarity (\(s_c\)) is computed. Each keyword belongs to one of several broad scientific categories (e.g., Stars, Galaxies, and Cosmology; see Appendix~\ref{subsec:eso_categories} for the full list). Category vectors are built similarly to keyword vectors, using the position of the first appearance of each category as a weight. 

The total matching score is defined as:
\[
s_{ij} = s_k + s_c,
\]
ranging from 0 (i.e., no keyword or category match) to 2 (i.e., perfect keyword, category, and order similarity). The total score serves as the metric to identify and rank reviewer-proposal pairs to assign ten reviewers to each proposal. In this work we utilize the keyword based similarity scores computed for P110 provided by ESO for evaluation as a baseline. 

\subsection{Topic Modeling with Latent Dirichlet Allocation}
LDA \citep{10.5555/944919.944937} is a generative probabilistic model that represents documents as mixtures of latent topics and each topic is described by a distribution of words. Probabilistic models such as LDA extend beyond a static, predefined set of keywords by describing both reviewers and proposals through shared conceptual topics. Formally, LDA assumes that each document $d$ is represented by a multinomial distribution over $K$ topics drawn from a Dirichlet prior with hyperparameter $\alpha$, where $K$ is an input. Each topic $k$ is itself a multinomial distribution of words, drawn from a Dirichlet prior with hyperparameter  $\beta$. The vector $\theta_d$ represents the topic proportions for document $d$, serving as its numerical representation. 

We determine the topic distributions for both reviewers and proposals using a jointly trained LDA model. In our preprocessing step, we concatenate all abstracts associated with a given reviewer into a single, comprehensive document (i.e., one document per reviewer) prior to model training. These reviewer-level documents are then combined with the individual proposal abstracts to create the final training corpus. By training the LDA model on this combined corpus, we directly infer the reviewer expertise vectors ($\theta_{r_j}$) and the proposal topic vectors ($\theta_{p_i}$) within the same latent topic space. We utilize the \textsc{gensim} library \citep{rehurek_lrec} for implementation. We preprocess the text by removing stop words (e.g., to, and, because, etc), as these high-frequency terms provide minimal semantic information and may skew the resulting topic distributions \citep{Angelov2020Top2VecDR}. The model was trained on the union of all queried reviewer publication abstracts and proposal abstracts to ensure a comprehensive vocabulary. Determining the appropriate number of topics, $K$, is a critical step as too few topics may overly generalize and too many may fragment meaningful concepts. While the ESO keywords are organized into nine broad categories, we experiment with the sensitivity of $K$ in Section~\ref{sec:ablation_results} to determine the optimal granularity for our topic model. The ALMA DPR system has adopted LDA to model reviewer expertise based on an investigator's previously submitted proposals \citep{carpenter_enhancing_2025}.

\subsection{Term Frequency-Inverse Document Frequency}
TF-IDF leverages the statistical word information of a set of documents to construct feature vectors that reflect the importance of terms within a single document relative to the entire set of documents \citep{luhn_statistical_1957,jones_statistical_nodate}. Formally, the TF–IDF weight for a term $t$ in a document $d$ is defined as:

$$\text{TF-IDF}(t, d) = \text{TF}(t, d) \times \text{IDF}(t)$$We utilize the scikit-learn implementation \citep{pedregosa_scikit-learn_2018}:$$\text{TF}(t, d) = f_{t, d}$$with $f_{t, d}$ denoting the number of times term $t$ appears in document $d$. The inverse document frequency down-weights terms that appear frequently across the corpus while up-weighting rare terms specific to a particular context. We utilize the smoothed $IDF$ formulation from the \texttt{scikit-learn} library, defined as:$$\text{IDF}(t) = \log \left( \frac{1 + N}{1+ n_t} \right) + 1$$where $N$ is the total number of documents in the corpus, and $n_t$ is the number of documents containing term $t$.

This method is utilized by the STScI Proposal Auto-Categorizer and Manager (PACMan) tool to represent reviewer expertise in its panel peer review process \citep{strolger_proposal_2017}. To achieve this, PACMan queries abstracts from NASA/ADS and computes the similarity between TF–IDF vector representations of reviewer publications and proposal content (specifically, the abstract and scientific justification) \citep{strolger_pacman2_2023}.

For each reviewer, the expertise vector is constructed by concatenating all retrieved abstracts into a single document, and proposal vectors are created from their respective abstracts. Both resulting documents are then vectorized using $\text{TF-IDF}$. We experiment with several hyperparameters to optimize the representation quality. Specifically, we vary the n-gram range to capture both unigrams and bigrams, allowing the model to incorporate two-word expressions (e.g., “stellar evolution”, “galaxy formation”, or "white dwarf") that convey more domain-specific meaning than separate individual words (e.g., "white", "dwarf"). Additionally, standard English  words (i.e., stop words) are removed to reduce noise (e.g., articles, prepositions, etc). 

\subsection{Transformer Representations}\label{sec:transformer_SI}
Transformer models extend beyond word-level statistics or topic-based approaches by capturing semantic relationships between words and phrases based on the surrounding context, enabling more sophisticated representations \citep{devlin2019bertpretrainingdeepbidirectional}. Leveraging direct semantic encoding via attention mechanisms \citep{attention_2017}, the Transformer architecture has been widely adapted for specific domains and tasks \citep{beltagy2019scibertpretrainedlanguagemodel, cohan_specter_2020}.

In this work, we evaluate two distinct Transformer models: the domain-specific $\text{\textsc{SPECTER2}}$ \citep{cohan_specter_2020} and the general-purpose $\text{\textsc{Sentence Transformer}}$ model ($\text{\textsc{all-distilroberta-v1}}$) \citep[an extension of][]{Sanh2019DistilBERTAD}.

First, the \textsc{SPECTER2} model is specifically designed for scientific document representation. It is initialized with \textsc{SciBERT} \citep{beltagy2019scibertpretrainedlanguagemodel} and fine-tuned using a citation-based contrastive loss, enforcing that papers citing each other have similar embedding representations. We utilize the $\text{\textsc{Sentence-Transformers}}$ framework \citep{reimers-2019-sentence-bert} because it is specifically fine-tuned using a contrastive objective to produce semantically meaningful, fixed-length embeddings from full sentences, unlike base models optimized for word-level prediction making them useful for similarity comparisons. The specific model we utilize, $\text{\textsc{all-distilroberta-v1}}$, is trained on over one billion sentence pairs, optimized to capture broad semantic similarity across diverse domains.

A primary challenge in utilizing Transformer methods for expertise modeling is the fixed input context window, typically limited to $\num{512}$ tokens, which is insufficient to embed a reviewer's full publication history simultaneously. To address this, we adopt a common two-stage aggregation approach. We independently encode each abstract in a reviewer’s history into a vector $\mathbf{p}_i$, obtaining a set of abstract vectors $\{\mathbf{p}_1, \dots, \mathbf{p}_N\}$. We then apply mean or max pooling across these vectors to produce a single, centroid-based expertise vector $\mathbf{r}_j$: $$\mathbf{r}_j = \frac{1}{N} \sum_{i=1}^N \mathbf{p}_i$$

In addition to encoder models that represent text and output vector embeddings, we evaluate a generative machine learning model using the \textsc{\textsc{GPT-4o mini}} model accessible via the OpenAI API \citep{openai2024gpt4ocard}. To implement this, we construct a structured system prompt where the LLM acts as an expert in astronomical time allocation (see Appendix~\ref{app:prompt}). We retrieve the reviewer publication histories via the NASA/ADS API to construct the context window for each evaluation. Deploying LLMs for pairwise comparisons at the scale of astronomical observatories presents significant cost challenges. To address this, we leverage prompt caching since a single reviewer must be compared against the \num{435} proposals, re-sending the reviewer's full publication history for every comparison is redundant. Finally, to mitigate the non-determinism of generative models and ensure scientific reproducibility, we explicitly set the sampling temperature to $T=0$ and fix the random seed (seed=$42$). Our framework leverages guard rails, context retrieval, and reproducibility constraints (e.g., harnesses \citep{yao2023reactsynergizingreasoningacting}). We deliberately select \textsc{\textsc{GPT-4o mini}} over reasoning models (e.g., the \textsc{GPT-5.5} series), as the latter utilize hidden chain-of-thought processes that currently restrict temperature control and introduce inherent non-determinism, rendering them unsuitable for strictly reproducible benchmarks. This configuration maximizes the stability of the outputs, ensuring that the relevance scores remain consistent across repeated experimental runs. 


\subsection{\textsc{GPT-4o mini} Prompt Design}
\label{app:prompt}

To estimate expertise using the \textsc{GPT-4o mini} model, we provided the system with the following zero-shot prompt. The model was instructed to output a scalar score between 0 and 100 based on the semantic alignment between the proposal abstract and the reviewer's publication history.

\begin{lstlisting}[caption={System prompt used for expertise scoring}, label={lst:gpt_prompt}]
You are an expert in assigning reviewers to proposals at astronomical observatories.
You want to make sure that the reviewers can give high quality and relevant reviews to the proposal they are assigned.

You are given the following input:
- "REVIEWER PAPERS", which is a selection of the most recent papers by the reviewer, containing the title and abstract of each paper.
- "NEW PROPOSAL", which contains the proposal abstract that is under consideration for assignment.

Your task is to assign a score (0-100) evaluating how well the NEW PROPOSAL matches the REVIEWER'S PAPERS.

Consider the following criteria:
1. The score should be based on the similarity between the proposal and the reviewer papers.
2. The score should be higher if the reviewer has more background knowledge and expertise in the proposal.
3. The score should be lower if the reviewer has less background knowledge and expertise in the proposal.

Scoring Scale: Assign any integer score from 0 to 100.
NOTE: Output ONLY the score. Use integer or float. Do not hallucinate.
\end{lstlisting}

\subsection{NDCG Definition for Self-reported Expertise Evaluation}\label{subsec:NDCG_SI}

To evaluate the similarity scores alignment with the reviewer self-reported labels (e.g., Expert, Intermediate, Non-expert), we utilized the Normalized Discounted Cumulative Gain (NDCG) metric. Unlike standard recall metrics, NDCG is rank-aware, explicitly rewarding algorithms that confidently place experts at the top of the assignment list.

NDCG is built upon Discounted Cumulative Gain (DCG), which measures a ranking based on its position and a graded relevance score. For example, for a single proposal, DCG is computed as:$$\text{DCG}_k = \sum_{j=1}^{k} \frac{\text{rel}_j}{\log_2(j + 1)}$$where $\text{rel}_j$ is the graded relevance score (i.e., self-reported expertise labels) of the result at position $j$, and $k$ is the total number of reviewers (e.g., \num{10} in ESO DPR). The formal definition for NDCG is:$$\text{NDCG}_k = \frac{\text{DCG}_k}{\text{IDCG}_k}$$where $\text{IDCG}_k$ is the maximum possible DCG calculated by placing all relevant reviewers in the ideal rank order. We utilize the scikit-learn implementation of $\text{NDCG}$ \citep{pedregosa_scikit-learn_2018}.

\section{Expertise Matrix Statistics and Ablation Studies}\label{sec:extended_results_SI}

\subsection{Expertise Matrices Statistics}\label{app:matrix_stats}
Table \ref{tab:distribution_stats} provides a statistical summary of the expertise matrices generated by each method. We report the sparsity (percentage of zeros) and the distribution quartiles to illustrate the varying density and scaling of the scores across different algorithms. All methods apart from the Keywords utilized up to \num{25} abstracts queried from NASA/ADS. 

\begin{table*}[h!]
\centering
\caption{Expertise Matrix Distribution Statistics}
\begin{tabular}{lccccccc}
\hline
Method  & \% zeros & min & 25th & median & 75th & max \\
\hline
Keywords  & 51.78\% & 0.00 & 0.00 & 0.00 & 0.43 & 1.00 \\
LDA$^\dagger$  & 67.93\% & 0.00 & 0.00 & 0.00 & 0.02 & 1.00 \\
TF--IDF  & 0.66\% & 0.00 & 0.03 & 0.04 & 0.06 & 0.50 \\ 
SPECTER2 (Mean pooling)  & 0.00\% & 0.75 & 0.91 &  0.92 & 0.94 & 0.98 \\
SentenceTransformer (Mean pooling)  & 0.00\% & -0.10 & 0.47 & 0.55 & 0.62 & 0.88 \\ 
GPT--4o mini   & 10.19\% & 0.00 & 0.25 & 0.45 & 0.75 & 0.95 \\ 
\hline
\multicolumn{7}{l}{\footnotesize $^\dagger$ Values below 0.01 for LDA are treated as zero.}
\end{tabular}
\label{tab:distribution_stats}
\end{table*}

\subsection{Ablation Results}\label{sec:ablation_results}
This section presents the full results of our ablation studies. Tables \ref{tab:ablation_1}, \ref{tab:ablation_2}, and \ref{tab:ablation_3} demonstrate the robustness of our retrieval performance under varying query parameters (timeframes and paper counts). Additionally, Table \ref{tab:k_topics} examines the impact of varying the topic count $K$ in the LDA algorithm, while Table \ref{tab:maxpooling} evaluates max pooling in the SPECTER2 and SentenceTransformer models.

We evaluate how different queries for publications from the NASA/ADS API affect retrieval performance. The API enables flexible querying along several dimensions, for example: (1) the maximum number of publications returned, (2) the publication time window, and (3) authorship position (e.g., first-author papers). Retrieving the correct publications for a reviewer is important as shorter time windows, smaller publication sets, and stricter authorship filters can reduce the amount and quality of information available for representing reviewer expertise.

Our baseline results use a query that retrieves \num{25} publications from the past five years, regardless of author position. To examine the sensitivity of the methods to changes in data retrieval, we evaluate three additional configurations: retrieving \num{50} publications from the last ten years, twelve publications from the last two years, and ten first-author publications from the last 5 years. These ablations allow us to assess how each representation method behaves under sparse versus more extensive publication histories.

Across all ablation settings, we observe a consistent trend: increasing the number of retrieved publications yields only marginal gains for all methods, whereas restricting the number of publications leads to degrading metrics. However, the magnitude of this sensitivity varies significantly by methodology. TF-IDF and SentenceTransformer demonstrate robustness, exhibiting only minor performance declines with fewer publications. In contrast, LDA and SPECTER2 are highly sensitive with performance decreasing when insufficient publications are queried. The keyword-based method remains static, as it is independent of the retrieval parameters. Note that we omit \textsc{GPT-4o mini} from this ablation due to cost constraints across multiple experiments ($>$\$300 USD per run). 

Table \ref{tab:k_topics} presents the results of our ablation of the sensitivity of the LDA model to the hyperparameter $K$ (number of topics). We observe a distinct performance peak at $K=50$, which yields the best median rank of \num{24.0} and Hit@25 of \num{0.503}. Lowering the number of topics decreases all metrics with $K=15$ yielding the lowest scores, while increasing $K$ to 75 reduces retrieval accuracy. Finally, Table~\ref{tab:maxpooling} shows that the Max pooling variants of SentenceTransformer and SPECTER2 exhibit the weakest overall performance. Both models recorded the lowest retrieval scores, yielding Median Ranks of \num{77.0} and \num{109.0}, respectively.

\begin{table*}[h!]
\centering
\caption{Ablation 1: Average rank-based retrieval performance across 435 proposals when querying for 12 papers in the last 2 years}
\begin{tabular}{lcccc}
\hline
Method & Median Rank $\downarrow$ & MRR $\uparrow$ & Hit@25 $\uparrow$ & Z-score $\uparrow$\\
\hline
Keywords & 9.0 \scriptsize{$\pm 2.5$} & 0.270 \scriptsize{$\pm 0.032$} & 0.703 \scriptsize{$\pm 0.044$} & 2.051 \scriptsize{$\pm 0.095$}\\
LDA (K = 50) & 42.0 \scriptsize{$\pm 6.0$}$^\dagger$ & 0.093 \scriptsize{$\pm 0.018$}$^\dagger$ & 0.395 \scriptsize{$\pm 0.046$}$^\dagger$ & 1.672 \scriptsize{$\pm 0.279$}$^\dagger$ \\
TF--IDF & \textbf{6.0} \scriptsize{$\pm 2.0$} & \textbf{0.340} \scriptsize{$\pm 0.035$}$^\dagger$ & \textbf{0.720} \scriptsize{$\pm 0.043$} & \textbf{3.583} \scriptsize{$\pm 0.343$}$^\dagger$ \\
SPECTER2 (Mean pooling) & 15.0 \scriptsize{$\pm 5.5$}$^\dagger$ & 0.256 \scriptsize{$\pm 0.034$} & 0.566 \scriptsize{$\pm 0.046$}$^\dagger$ & 0.778 \scriptsize{$\pm 0.111$}$^\dagger$ \\
SentenceTransformer (Mean pooling) & 9.0 \scriptsize{$\pm 2.5$}$^*$ & 0.304 \scriptsize{$\pm 0.035$} & 0.662 \scriptsize{$\pm 0.045$} & 1.109 \scriptsize{$\pm 0.122$}$^\dagger$ \\
\hline
\end{tabular}
\label{tab:ablation_1}
\end{table*}

\begin{table*}[h!]
\centering
\caption{Ablation 2: Average rank-based retrieval performance across 435 proposals when querying for 50 papers in the last 10 years}
\begin{tabular}{lcccc}
\hline
Method & Median Rank $\downarrow$ & MRR $\uparrow$ & Hit@25 $\uparrow$ & Z-score $\uparrow$\\
\hline
Keywords & 9.0 \scriptsize{$\pm 1.0$} & 0.275 \scriptsize{$\pm 0.033$} & 0.701 \scriptsize{$\pm 0.043$} & 2.051 \scriptsize{$\pm 0.096$}\\
LDA (K = 50) & 19.0 \scriptsize{$\pm 4.0$ \dag} & 0.166 \scriptsize{$\pm 0.025$ \dag} & 0.568 \scriptsize{$\pm 0.046$ \dag} & 2.540 \scriptsize{$\pm 0.238$ \dag} \\
TF--IDF & \textbf{4.0} \scriptsize{$\pm 1.0$ \dag} & \textbf{0.432} \scriptsize{$\pm 0.038$ \dag} & \textbf{0.807} \scriptsize{$\pm 0.037$ \dag} & \textbf{3.962} \scriptsize{$\pm 0.298$ \dag} \\
SPECTER2 (Mean pooling) & 9.0 \scriptsize{$\pm 3.0$ *} & 0.294 \scriptsize{$\pm 0.034$} & 0.634 \scriptsize{$\pm 0.046$ *} & 0.884 \scriptsize{$\pm 0.114$ \dag} \\
SentenceTransformer (Mean pooling) & 6.0 \scriptsize{$\pm 1.5$} & 0.352 \scriptsize{$\pm 0.037$ \dag} & 0.720 \scriptsize{$\pm 0.041$} & 1.377 \scriptsize{$\pm 0.122$ \dag} \\
\hline
\end{tabular}
\label{tab:ablation_2}
\end{table*}

\begin{table*}[h!]
\centering
\caption{Ablation 3: Average rank-based retrieval performance across 435 proposals when querying for 10 first author papers in the last 5 years}
\begin{tabular}{lcccc}
\hline
Method & Median Rank $\downarrow$ & MRR $\uparrow$ & Hit@25 $\uparrow$ & Z-score $\uparrow$\\
\hline
Keywords & 9.0 \scriptsize{$\pm 2.5$} & 0.270 \scriptsize{$\pm 0.032$} & \textbf{0.703} \scriptsize{$\pm 0.044$} & 2.051 \scriptsize{$\pm 0.095$}\\
LDA (K = 50) & 62.0 \scriptsize{$\pm 14.5$}$^\dagger$ & 0.091 \scriptsize{$\pm 0.019$}$^\dagger$ & 0.317 \scriptsize{$\pm 0.043$}$^\dagger$ & 1.338 \scriptsize{$\pm 0.220$}$^\dagger$ \\
TF--IDF & \textbf{5.0} \scriptsize{$\pm 1.5$} & \textbf{0.424} \scriptsize{$\pm 0.041$}$^\dagger$ & 0.678 \scriptsize{$\pm 0.043$} & \textbf{4.643} \scriptsize{$\pm 0.490$}$^\dagger$ \\
SPECTER2 (Mean pooling) & 29.0 \scriptsize{$\pm 15.0$}$^\dagger$ & 0.297 \scriptsize{$\pm 0.038$} & 0.494 \scriptsize{$\pm 0.047$}$^\dagger$ & 0.644 \scriptsize{$\pm 0.123$}$^\dagger$ \\
SentenceTransformer (Mean pooling) & 12.0 \scriptsize{$\pm 4.5$}$^\dagger$ & 0.356 \scriptsize{$\pm 0.040$}$^*$ & 0.575 \scriptsize{$\pm 0.047$}$^\dagger$ & 0.736 \scriptsize{$\pm 0.129$}$^\dagger$ \\
\hline
\end{tabular}
\label{tab:ablation_3}
\end{table*}

\begin{table*}[h!]
\centering
\caption{Ablation 4: Average rank-based retrieval performance across 435 proposals when querying 25 publications and varying the number of K topics in the LDA algorithm}
\begin{tabular}{lcccc}
\hline
Method & Median Rank & MRR & Hit@25 & Z-score\\
\hline
LDA (K = 15) & 30.0 \scriptsize{$\pm 5.5$} & 0.106 \scriptsize{$\pm 0.020$} & 0.453 \scriptsize{$\pm 0.048$} & 1.642 \scriptsize{$\pm 0.123$} \\
LDA (K = 25) & 29.0 \scriptsize{$\pm 5.0$} & 0.118 \scriptsize{$\pm 0.020$} & 0.467 \scriptsize{$\pm 0.046$} & 1.911 \scriptsize{$\pm 0.183$} \\
LDA (K = 50) & \textbf{24.0} \scriptsize{$\pm 6.0$} & \textbf{0.148} \scriptsize{$\pm 0.024$} & \textbf{0.503} \scriptsize{$\pm 0.047$} & 2.114 \scriptsize{$\pm 0.208$} \\
LDA (K = 75) & 28.0 \scriptsize{$\pm 6.5$} & 0.143 \scriptsize{$\pm 0.025$} & 0.483 \scriptsize{$\pm 0.046$} & \textbf{2.266} \scriptsize{$\pm 0.363$} \\ 
\hline
\end{tabular}
\label{tab:k_topics}
\end{table*}

\begin{table*}[h!]
\centering
\caption{Ablation 5: Average rank-based retrieval performance across 435 proposals when employing Max pooling for Transformer models}
\resizebox{\textwidth}{!}{
\begin{tabular}{lcccc}
\hline
Method & Median Rank $\downarrow$ & MRR $\uparrow$ & Hit@25 $\uparrow$ & Z-score$\uparrow$ \\
\hline
SPECTER2 (Max pooling) & 109.0 \scriptsize{$\pm 23.5$} & 0.035 \scriptsize{$\pm 0.013$} & 0.110 \scriptsize{$\pm 0.030$} & 0.256 \scriptsize{$\pm 0.101$} \\
SentenceTransformer(Max pooling) & 77.0 \scriptsize{$\pm 20.5$ } & 0.071 \scriptsize{$\pm 0.017$} & 0.290 \scriptsize{$\pm 0.044$} & 0.628 \scriptsize{$\pm 0.121$} \\ 
\hline
\end{tabular}
}
\label{tab:maxpooling}
\end{table*}

\section{Methodological Limitations}\label{sec:limitations_SI}
A key limitation of relying on the NASA/ADS API for automatically querying a researcher's publications is the lack of author name disambiguation. As \citet{olivo2025practicalauthordisambiguationmetadata} note, an estimated 52\% of name groupings (i.e., first initial and last name) in NASA/ADS contain ambiguity, which can introduce noise into the expertise representations when researchers share similar names \citep{strolger_proposal_2017}. To mitigate this in a production system, observatories could adopt the approach used by platforms like OpenReview, where reviewers manually select their representative publications. This ensures accurate, up-to-date profiles while eliminating the algorithmic risks associated with querying digital library APIs. 
We verified the data completeness for the \num{379} researchers in our dataset by querying their last \num{25} publications over a five-year window. Despite the inclusion of student reviewers in the Distributed Peer Review process, only two individuals returned zero records. We resolved one case by extending the search window to ten years, and noted the other possessed only unrefereed publications. In a live production environment, the expertise of early-career reviewers returning zero publications could be accurately represented by utilizing the text of the proposal they submitted to the current cycle, guaranteeing a baseline expertise vector \citep{carpenter_enhancing_2025}. 
Finally, utilizing a generative Large Language Model (e.g., \textsc{GPT-4o mini}) introduces non-trivial computational costs ($\approx$\$300-500 USD per experimental run in our benchmark). Furthermore, these models rely on probabilistic generation, which inherently limits the strict deterministic reproducibility required for a standardized, fair peer review pipeline.
\end{document}